\documentclass[
    ,final            % use final for the camera ready runs
,sort&compress
  ]
  {aipproc}

\layoutstyle{8x11double}

\newcommand{\nodata}{\ldots}

\newcommand{\mc}{\multicolumn}
\newcommand{\pasp}{PASP}
\newcommand{\araa}{ARA\&A}
\newcommand{\aplett}{ApL}
\newcommand{\apj}{ApJ}
\newcommand{\apjs}{ApJS}
\let\apjl\apj
\newcommand{\aj}{AJ}
\newcommand{\apss}{Ap\&SS}
\newcommand{\aap}{A\&A}

\newcommand{\pre}{Phys.\ Rev.\ E}
\newcommand{\physrep}{Phys.\ Rep.}

\newcommand{\mnras}{MNRAS}
\newcommand{\nat}{Nature}

\newcommand\phn{\phantom{0}}%

\newcommand{\rosat}{\textit{ROSAT}}
\newcommand{\xmm}{\textit{XMM-Newton}}
\newcommand{\chandra}{\textit{Chandra}}

\def\gsim{\mathrel{\hbox{\rlap{\lower.55ex \hbox {$\sim$}}
                   \kern-.3em \raise.4ex \hbox{$>$}}}}
\def\lsim{\mathrel{\hbox{\rlap{\lower.55ex \hbox {$\sim$}}
                   \kern-.3em \raise.4ex \hbox{$<$}}}}
\newcommand{\expnt}[2]{\ensuremath{#1 \times 10^{#2}}}   % scientific notation

\newcommand{\Rxjw}{RX~J1856.5$-$3754}
\newcommand{\Rbs}{RX~1308.6+2127}
\newcommand{\Rxjk}{RX~J0720.4$-$3125}
\newcommand{\Rxj}{RX~J1605.3+3249}

\newcommand{\rxjw}{RX~J1856}
\newcommand{\rbs}{RX~1308}
\newcommand{\rxjk}{RX~J0720}
\newcommand{\rxj}{RX~J1605}

\begin{document}

\title{Nearby, Thermally Emitting Neutron Stars}

\classification{97.60.Jd, 97.60.Gb 	}
\keywords      {neutron stars, pulsars}

\author{David~L.~Kaplan}{
  address={Pappalardo Fellow, Kavli Institute for Astrophysics and Space Research and
  Department of Physics,
  Massachusetts Institute of Technology, 77 Massachusetts Ave, Room
  37-664H, Cambridge, MA 02139}
}

%\author{M.~H.~van~Kerkwijk}{
%  address={Department of Astronomy \& Astrophysics,
%           University of Toronto,
%           60 Saint George Street,
%           Toronto, Ontario~~M5S 3H8,
%           Canada}
%}
\begin{abstract}
We describe a sample of thermally emitting neutron stars discovered in
the \rosat\ All-Sky Survey.  We discuss the basic observational
properties of these objects and conclude that they are nearby,
middle-aged pulsars with moderate magnetic fields that we see through
their cooling radiation. While these objects are potentially very
useful as probes of matter at very high densities and magnetic fields,
our lack of understanding of their surface emission limits their
current utility.  We discuss this and other outstanding problems: the
spectral evolution of one sources and the relation of this population
to the overall pulsar population.
\end{abstract}

\maketitle

\section{Introduction}
While almost 2000 isolated neutron stars have now been discovered as
radio pulsars, the total number in the Galaxy is much larger.  Radio
pulsars emit pulsations for $\sim10^7$~yr and are visible due to radio
beams that subtend 1--10\% of the sky, so the total number of neutron
stars of all ages just in the local region of the Galaxy (where radio pulsars are
detectable) should be $\gsim 10^6$ \citep[e.g.,][]{vml+04,fgk06}.

Are those neutron stars not detectable as radio pulsars objects
invisible, or is there some chance of observing them (there is always
the exception of Geminga, where we see no radio emission but is
otherwise a standard pulsar)?  For years astronomers have proposed
that a large fraction of these objects would be visible through one of
two mechanisms: accretion \citep{ors70,tc91,bm93} or cooling
\citep{ttzc00}.  The first mechanism could revive old, dead pulsars,
while the second would primarily work for younger sources.  Both
mechanisms, however, make the neutron stars visible in the soft X-ray
regime, not in the radio regime that had dominated the study of
neutron stars.

These  neutron stars should be identifiable by \citep{tc91}:
\begin{enumerate}
\item Largely thermal emission peaking in the soft X-ray or far-UV
  band, requiring small hydrogen column densities to remain visible
\item The absence of bright optical counterparts
\item Significant ($\gsim 0.1\mbox{ arcsec yr}^{-1}$) proper motions
\item Preferred locations in the Galactic plane
\end{enumerate}
The first two criteria relate to the spectra of the neutron stars, and
serve to rule out the active galaxies and stars that dominate X-ray
surveys \citep{hg88}.  The third criterion reflects the proximity of
the sources (with maximum distances of $\sim 1$~kpc) and the large
space velocities of known neutron stars \citep{hllk05,fgk06}.  The
final criterion comes from the Galactic nature of the sources, and is
similar to the distribution of radio pulsars.  

%% The second criterion relates to many classes of neutron stars, not
%% just the accreting/cooling sources discussed here.  In fact, due to
%% their small sizes and hot temperatures, neutron stars that are
%% detectable in bands outside the radio regime generally have very high
%% ratios of X-ray to optical flux.  For thermal sources, this is
%% approximately
%% \begin{equation}
%% \frac{L_{\rm X}}{L_{\rm opt}} \sim 10^{5.5+3\log(kT/100\mbox{ eV})}
%% \end{equation}
%% \citep{ttzc00,rfbm03}.  This compares to stars values of
%% $10^{-3}$--$10^{-2}$ for stars \citep{kc00} and 0.1--10 for active
%% galaxies \citep{bah+01}.  Only
%% white dwarfs and X-ray binaries (compact objects, like neutron stars)
%% can come close, with ratios of 10--1000 \citep{hg88}.  

\subsection{The Legacy of \rosat}
While some had anticipated up to 5,000 objects discovered through soft
X-ray surveys \citep{tc91,bm93}, the \rosat\ All-Sky Survey (RASS)
discovered just over half a dozen\footnote{The difference is largely
  from poor assumptions about the pulsar velocity distribution 
 \citep{nt99,ttzc00} and the effects of magnetic fields
\citep{is75,pnr+03}.} (as of 2007) nearby cooling neutron stars (see
Tab.~\ref{tab:sum}).  These neutron stars are then all the more
valuable because of their rarity.

The first such source to be discovered was \Rxjw\ (hereafter \rxjw).
It was originally identified serendipitously as a soft, bright X-ray
source with no obvious optical counterpart \citep{wwn96}.  Its
location in front of the R~CrA molecular cloud meant that it had to be
nearby ($\lsim 200$~pc \citep{kh98}) and hence small (not a white
dwarf or anything larger)---otherwise the X-ray emission would have
been absorbed\footnote{As it turned out, this argument was false, as
observations of more distant stars did not have high extinction
\citep{vkk01}.  Nonetheless, \rxjw\ is quite close
\citep{kvka02,wl02}.}.  Confirmation of its nature came with the
discovery of a very faint ($m_B\approx 25.8$~mag), blue optical
counterpart \citep{wm97}.

\begin{table}
\setlength{\tabcolsep}{4pt}
{%\footnotesize
%\small
\begin{tabular}{cccccccccccccc}
\hline
\tablehead{1}{c}{c}{RX J} & \mc{3}{c}{\textbf{Spin}\tablenote{We
    give the spin-period, period derivative, and rms pulsed fraction.}
} & & \mc{5}{c}{\textbf{Spectrum}\tablenote{We give the hydrogen
    column density, blackbody temperature, \xmm\ EPIC-pn count-rate, central energy of any
    absorption features in the X-ray spectrum, and $B$-band magnitude.  We also give energies
    of any secondary lines, if known.  Note that all spectral
    estimates are covariant, and that the details of the X-ray
    absorption depend on the specific model.}
}  &&
\mc{2}{c}{\textbf{Astrometry}\tablenote{We give the proper motion and
    distance, using the parallax if known, else that inferred from \citet{pph+07}.}}
 & \tablehead{1}{c}{c}{References}

\\ \cline{2-4} \cline{6-10}
 \cline{12-13}
 & \tablehead{1}{c}{b}{$\mathbf{P}$} &
\tablehead{1}{c}{b}{$\mathbf{\dot P}$} & \tablehead{1}{c}{b}{PF} &&
\tablehead{1}{c}{b}{$\mathbf{N_{H,20}}$} &
\tablehead{1}{c}{b}{$\mathbf{kT}$} &
\tablehead{1}{c}{b}{PN}&\tablehead{1}{c}{b}{$\mathbf{E_{\rm abs}}$}&
\tablehead{1}{c}{b}{$\mathbf{m_B}$} 
&&
\tablehead{1}{c}{b}{$\mathbf{\mu}$} & \tablehead{1}{c}{b}{d} \\
& \tablehead{1}{c}{b}{(s)} &
\tablehead{1}{c}{b}{($\mathbf{10^{-14}}$)} &
\tablehead{1}{c}{b}{(\%)}& 
& \tablehead{1}{c}{b}{($\mathbf{\mbox{cm}^{-2}}$)} &
\tablehead{1}{c}{b}{(keV)} &
\tablehead{1}{c}{b}{($\mathbf{s^{-1}}$)}
&\tablehead{1}{c}{b}{(keV)}&\tablehead{1}{c}{b}{(mag)}&&
\tablehead{1}{c}{b}{($\mathbf{\mbox{mas yr}^{-1}}$)} &
\tablehead{1}{c}{b}{(pc)} 
 \\ \hline
1856.5$-$3754 & \phn7.06 & \nodata & \phn1  && 0.8 & \phn62 &8.3 &
\nodata &25.2&& 333 & 160 & \citenum{kvka02,bhn+03,tm07,vkk01,vkk07}\\
0720.4$-$3125\tablenote{The spectral parameters are average quantities.} & \phn8.39 & \phn7 & 11 && 1.0 & \phn87 &7.6 &  0.3 &26.6 &&\phn97& 360& \citenum{kvkm+03,mzh03,kvk05,kvka07,vkkpm07,hztb04}\\
1605.3+3249   & \nodata & \nodata &  $<3$  && 0.8 & \phn93 &5.6 &
 0.5(0.6,0.8) &27.2 && 155 & 390 & \citenum{kkvk03,vkkd+04,haberl07,msh+05,zdlmt06}\\
1308.6+2127   & 10.31 & 11 & 18 && 1.8 & 102 &2.5 &
 0.2(0.4) &28.4\tablenote{Inferred from STIS 50CCD.} &&
200\tablenote{Based on Motch~et~al., these proceedings.} & \nodata& 
\citenum{kkvk02,kvk05b,hsh+03,shhm05,shhm07}\\
2143.0+0654 & \phn9.44 & \nodata & \phn4 && 3.6 & 102 &2.0 &
 0.7 &$>26$\tablenote{Inferred from $V$ and $r^{\prime}$ bands.} &&
\nodata & 430 & \citenum{zct+01,zct+05,rtj+07}\\
0806.4$-$4123 & 11.37 & \nodata & \phn6 && 1.1 & \phn92 &1.8 &  0.3(0.6)
&$>24$&&\nodata & 250 & \citenum{hmz+04,haberl07}\\
0420.0$-$5022 & \phn3.45 & \nodata & 17 && 2.1 & \phn45 &0.2 & 0.3 
&26.6 &&\nodata & 345 & \citenum{hmz+04,haberl07}\\
\hline
%% \multicolumn{14}{p{0.95\textwidth}}{References: 
%% 1: \citet{kvka02},
%% 2: \citet{bhn+03}, 
%% 3: \citet{tm07},
%% 4: \citet{vkk01}, 
%% 5: \citet{vkk07},
%% 6: \citet{kvkm+03}, 
%% 7: \citet{mzh03}, 
%% 8: \citet{kvk05}, 
%% 9: \citet{kvka07}, 
%% 10: \citet{vkkpm07}, 
%% 11: \citet{hztb04},
%% 10: \citet{kkvk03}, 
%% 11: \citet{vkkd+04}, 
%% 15: \citet{haberl07}, 
%% 17: \citet{msh+05},
%% 12: \citet{kkvk02}, 
%% 13: \citet{kvk05b}, 
%% 14: \citet{hsh+03}, 
%% 16: \citet{shhm05}
%% 17: \citet{zct+01}, 
%% 18: \citet{zct+05}, 
%% 19: \citet{hmz+04}, 
%% 20: \citet{zdlmt06}, 
%% 21: \citet{rtj+07}, 
%% 23: \citetext{Motch~et~al.\ these proceedings}, 
%% 24: \citet{shhm07}.
%% }\\
\end{tabular}}
\caption{Observed Properties of the Seven Isolated Neutron
  Stars\label{tab:sum}}
\end{table}

Since then, six other similar sources have been identified with
considerable effort \citep{hmb+97,hmp98,shs+99,mhz+99,hpm99,zct+01}.
While a variety of names and acronyms exist for these objects, we call
them simply ``Isolated Neutron Stars'' (INS).  Identification of
additional sources that may still be present \citep{rfbm03} in the
\rosat\ Bright Sources Catalog (containing $\approx 18000$ sources
with $>0.05\mbox{ counts s}^{-1}$ with the Position-Sensitive
Proportional Counter, or PSPC; \citep{rbs}) is extremely difficult
given the poor positional accuracy of the PSPC.  We will return to
this later.

\section{Observed Properties}
We summarize the properties of the 7 confirmed INS in
Table~\ref{tab:sum}: also see previous reviews such as
\citet{haberl07}, that give additional observational details.  The
basic properties largely bear out the expectations of \citet{tc91} but
with much smaller numbers: they are soft X-ray sources with faint
optical counterparts and significant proper motions.  However, there
are too few of them to see a concentration in the Galactic plane, and
they may actually reflect a more local population that was born in the
Gould Belt \citep{ptp+05}.

\begin{figure}[t]
\hbox{\includegraphics[width=0.32\textwidth]{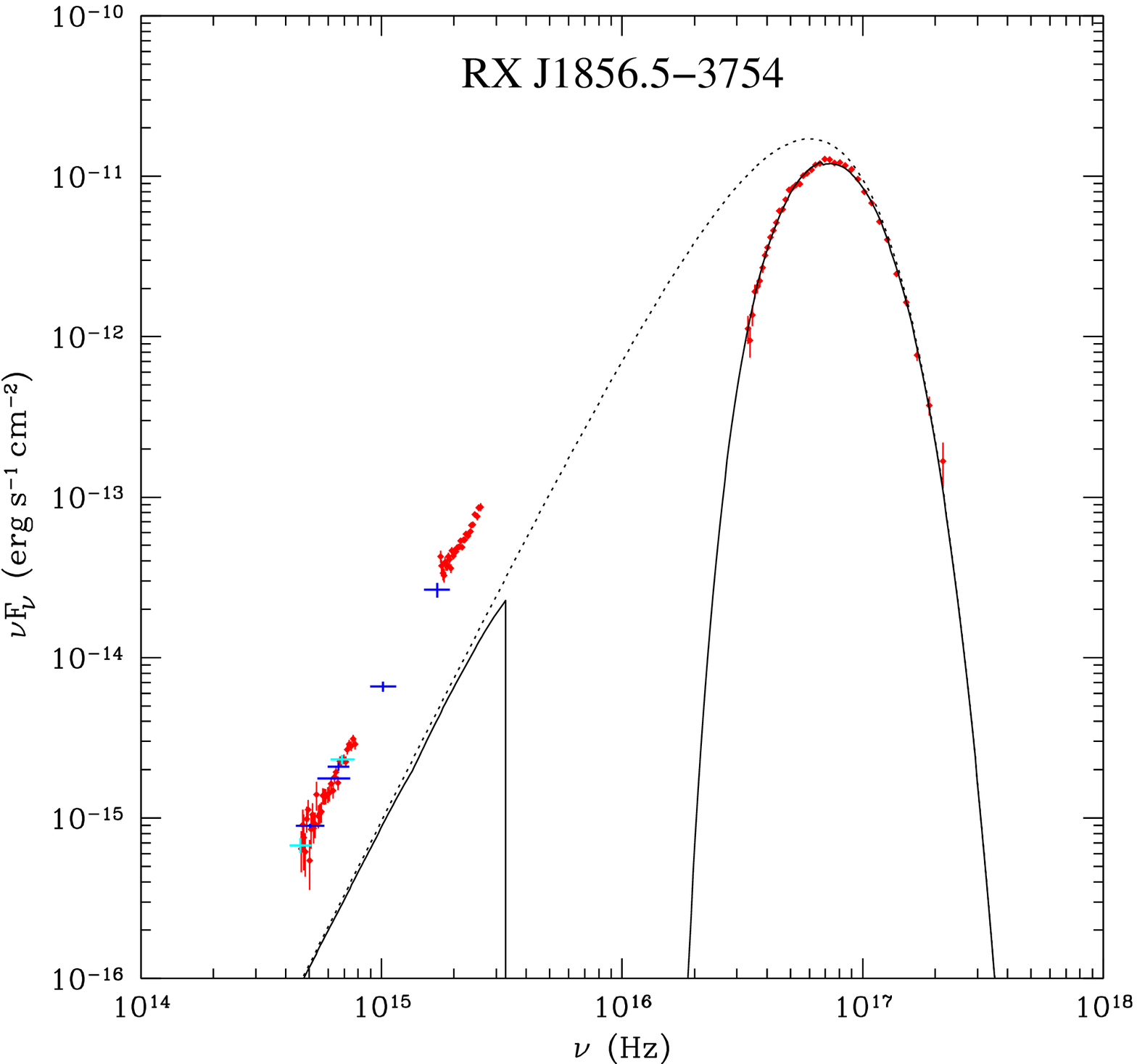}\includegraphics[width=0.32\textwidth]{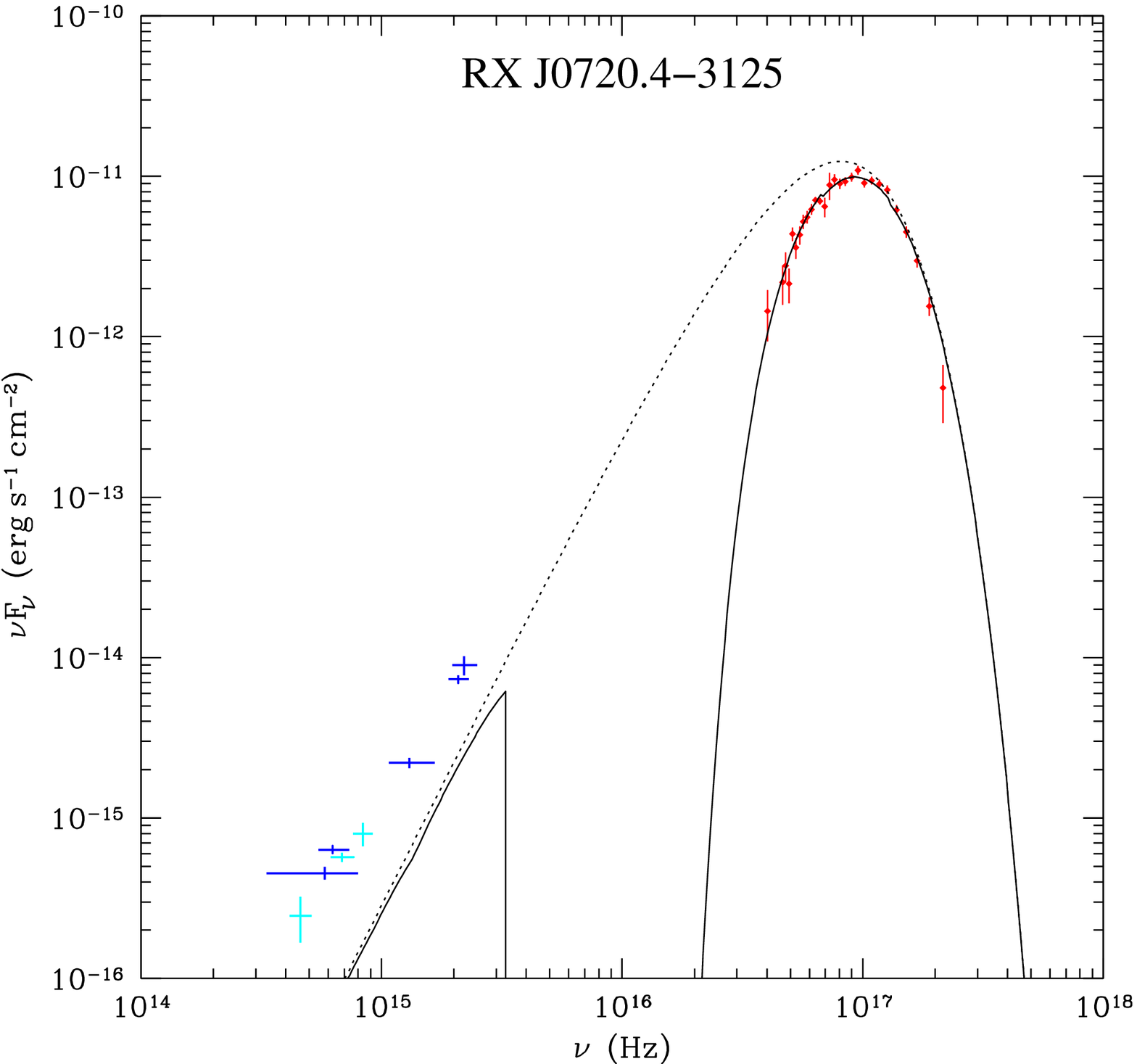}\includegraphics[width=0.32\textwidth]{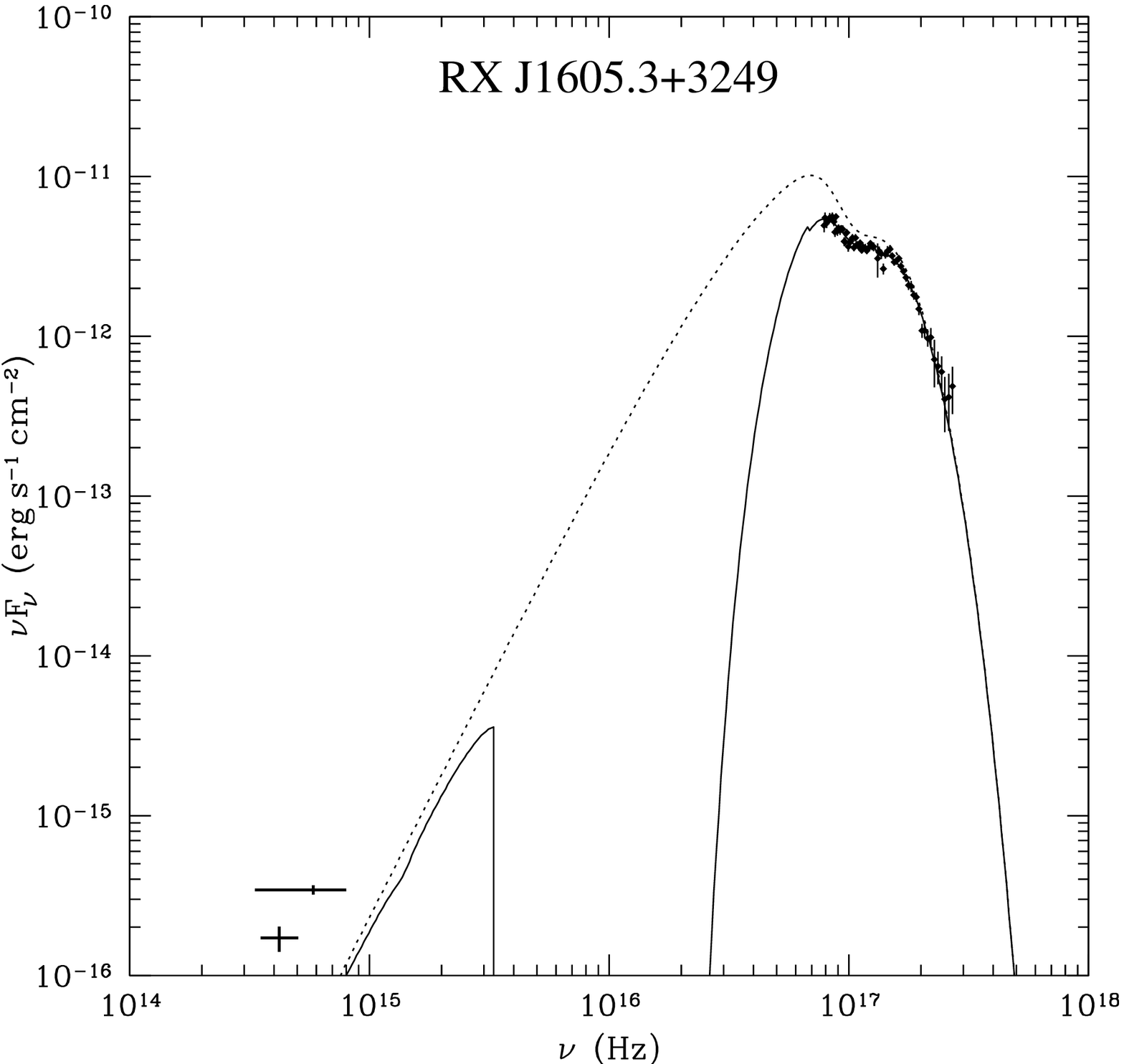}}
\caption{Spectral energy distributions for \rxjw\ (left), \rxjk\
  (center; with the X-ray spectrum from before the appearance of an
  absorption feature), and \rxj\ (right) .  For the first two, the X-ray points are from \chandra/LETG
  spectra \citep{bhn+03,kvkm+03}, and for \rxj\ the X-ray points are from
  \xmm/RGS \citep{vkkd+04}  The optical data are from a
  combination of ground-based  and space-based
   observations \citep{vkk01,kkvk03}.  The solid curves
  represent the best-fit black-body models to the X-ray data (with an
  absorption feature in the case of \rxj); the
  dotted curves are the same model without interstellar extinction.}
\label{fig:sed}
\end{figure}

%\paragraph{X-ray Spectra}
The X-ray spectra of the INS are reasonably close to soft black bodies
with $kT$ in the range of 40 to 100~eV \citep{bhn+03} attenuated by a
small amount of interstellar absorption (Fig.~\ref{fig:sed}).  There is no
sign of any non-thermal X-ray power-law such as those seen in the
spectra of most radio pulsars.  Most of the INS, however, have spectra
that appear with significant  absorption features at low energies (Fig.~\ref{fig:sed})
\citep{hsh+03,vkkd+04,hztb04,hmz+04,zct+05}.  Typically, the
absorption is modeled as one or more broad absorption lines, although
the line shapes are complex and phase-dependent, and as we obtain
increasing amounts of data our initial fits are no longer sufficient.  All but
one of the INS have hints of absorption in their spectra, and several
\citep{haberl07,shhm07} may even have absorption at multiple energies.
\citet{pph+07} used the measured column densities, along with a map of
the local interstellar medium, to infer distances to the INS.  They
found the objects were nearby ($\lsim 500$~pc, generally), although we
note that the it can be difficult to establish a reliable column
density when the intrinsic form of the spectrum is not well known
\citep[e.g.,][]{dvk06}.

\begin{figure}
\includegraphics[width=\hsize]{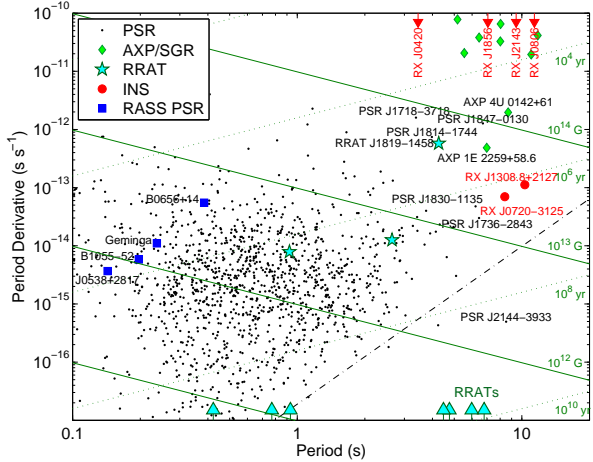}
\caption{A portion of the $P$-$\dot P$ diagram, showing radio pulsars (points; \citep{mhth05}),
magnetars (diamonds), and the newly-discovered rotating radio
transients (RRATs \citep{mll+06}; stars and arrows at bottom); selected
objects are labeled, and the pulsars with $>0.05\mbox{ s}^{-1}$ in the
RASS are shown as the squares.  Also shown are the  INS: RX~J0720 and RX~J1308
by filled circles, and the others by arrows at the top (since $\dot P$ is
unknown).  The diagonal lines show loci of constant dipole magnetic
field (solid) and spin-down age (dotted),  while the
dot-dashed line is an approximate ``death-line'' for pulsar activity.
The similarity in spin properties between the INS and some of the
RRATs was noted by \citet{mll+06} and \citet{ptp06}.}
\label{fig:ppdot}
\end{figure}

%\paragraph{Timing}
All but one of the INS show gentle, largely sinusoidal X-ray
pulsations, although some may be double-peaked \citep{shhm05}.  The
periods are all tightly clustered (compared to radio pulsars) between
3 and 11~s (Fig.~\ref{fig:ppdot}), and the pulsed fractions vary from
$\sim 1$\% to almost 20\%.  Through repeated X-ray observations we
have been able to establish reliable, phase-connected timing solutions
for two of the INS, \Rxjk\ (hereafter \rxjk) and \Rbs\ (hereafter
\rbs), and we find spin-downs of $\sim 10^{-13}\mbox{ s s}^{-1}$.
With the period derivatives, we can calculate the usual pulsar
quantities \citep{lk04}: dipolar magnetic field $B_{\rm dip}=2.4$ and
$3.4\times10^{13}$~G, characteristic age $\tau=1.9$ and 1.5~Myr, and
spin-down energy loss rate $\dot E=4.7$ and $\expnt{4.0}{30}\mbox{ erg
s}^{-1}$ for \rxjk\ and \rbs, respectively \citep{kvk05,kvk05b}.

%\paragraph{Optical \& Ultraviolet Spectra}
In the optical and ultraviolet, deep searches have revealed confirmed
counterparts to four of the INS \citep{wm97,kvk98,mh98,kkvk02,kkvk03},
and a possible counterpart to a fifth \citep{hmz+04}.  In keeping with
the identification of these objects as neutron stars, the optical
counterparts are quite faint (far fainter than any possible companion
might be), with X-ray to optical flux ratios of $\sim 10^4$.  However,
the optical fluxes generally lie a factor of $\sim 10$ above the
extrapolation of the X-ray spectrum (Fig.~\ref{fig:sed}): the so
called ``optical excess.''  In the best studied case, \rxjw, the
optical/UV spectrum is consistent with the slope of a Rayleigh-Jeans
tail: $F_{\nu}\propto \nu^2$ \citep{vkk01}.  In the other cases we do
not have nearly as much data and the inferences are consequently less
certain.  For \rxjk, the optical/UV spectrum is close to the slope of
a Rayleigh-Jeans tail, but there are indications that it deviates
\citep{kvkm+03,mzh03}.  For \Rxj\ (hereafter \rxj), the data are even
sparser, but it might have even larger deviations from a
Rayleigh-Jeans tail \citep{msh+05}.

%\paragraph{H$\alpha$ Nebulae}
An elongated H$\alpha$ nebula surrounding \rxjw\ was found by \citet{vkk01b}.
The nature of this nebula is not entirely clear, but it is likely a
bowshock  formed by the interaction between the neutron star's
energetic wind and the interstellar medium through which it is
traveling supersonically \citep[e.g.,][]{cc02}.  In this model, we can
infer $\dot E$ for \rxjw\ even though $\dot P$ has not been measured,
and find $\dot E\gsim \expnt{1.2}{33}d_{160}^3\mbox{ erg s}^{-1}$
(where the distance  is $160d_{160}$~pc).  \citet{msh+05}
discovered some evidence of H$\alpha$ emission from the position of
\rxj, but it has yet to be confirmed.

%\paragraph{Astrometry}
Proper motions have now been measured for four of the INS \citep[][and
Motch~et~al., these proceedings]{kvka02,mzh03,msh+05,kvka07}.  The
implied transverse velocities are generally $\sim 200\mbox{ km
s}^{-1}$, consistent with the pulsar population.  Tracing
the proper motions back, one finds that these INS are consistent with
being born in nearby OB associations $\sim 1$~Myr ago \citep{wal01}.
We have also been able to measure parallaxes to two of the INS.  The
distances are largely consistent with those estimated by
\citet{pph+07}.

%\paragraph{Radio \& Infrared}
In the radio, there are no confirmed detections of any INS as
a steady radio source or via pulsations
\citep{kkvk02,kkvk03,kvkm+03,johnston03}.  Motivated by the similar
location in the $P-\dot P$ diagram of the INS and the Rotating RAdio
Transients (RRATs; \citep{mll+06}), there have also been searches for
sporadic ratio bursts from the INS, and these too have been negative
\citetext{see Burgay,  Kondratiev in these 
  proceedings}.  Similarly, searches for counterparts in the
near-infrared (where one might see emission from a faint companion or
an accretion disk) have not been successful \citep{lcmpi07}.

%% \begin{figure}
%% \includegraphics[width=\hsize]{inslall.eps}
%% \caption{Calibrated \chandra\ LETG spectra of two INS, from \citet{vkkpm07}.  The spectrum of \rxjk\
%%    from 2000 (red open circles) is well described by an absorbed
%%    black-body (solid red curve).  The average post-2003 spectrum
%%    (magenta filled circles) could be a superposition of spectra from
%%    unchanged and changed parts of the surface.  As an indication, a
%%    close to maximum contribution by the unchanged spectrum is
%%    indicated (magenta, dotted line).  The residual (blue, wiggly
%%    curve, offset down by 0.1 unit for clarity) is remarkably similar
%%    to the spectrum of \rbs\ (green squares, scaled by a factor 1.2).
%%    For all spectra, the 40--44\,\AA\ range is less reliable, because
%%    of strong instrumental carbon absorption.}
%% \label{fig:letg}
%% \end{figure}

\section{ Inferences \&\ Energy Sources}
From the general properties of the INS discussed above, it seems clear
that the INS are reasonably young ($\sim 1$~Myr) isolated neutron
stars.  The X-ray emission that we see is plausibly thermal: the
values of $\dot E$ that we measure or infer are all quite low,
compared to X-ray luminosities of $\sim \expnt{3}{32}d_{360}^2\mbox{
  erg s}^{-1}$ (for \rxjk\ \citep{kvkm+03}, where we normalize to a
distance of $360d_{360}$~pc).  In contrast to radio pulsars (and other
rotation-powered objects such as Geminga), which have $L_{\rm X}\sim
10^{-3} \dot E$ \citep{bt97,pccm02} and where rotational energy powers
the majority of the X-ray emission (some may be thermal), \rxjk\ has $L_{\rm X}/\dot E\sim
60$.  Therefore, for \rxjk\ rotation \textit{cannot} power the X-ray
emission, and we must resort to other mechanisms.  While
 the value of $\dot E$ inferred for \rxjw\ from the H$\alpha$
nebula is of a different order than those for \rxjk\ and \rbs, it is
still much less than what would be necessary to contribute
significantly to the X-ray emission.  However, unlike for \rxjk\ where
the age inferred from timing is close to the kinematic age, using
the inferred $\dot E$ for \rxjw, its spin period, and kinematic age it
is hard to form a consistent picture, something we are investigating.

The remaining energy sources that we must consider are accretion,
magnetic fields, and thermal energy, all of which are seen in other
types of neutron stars.  

%\paragraph{Accretion}
Models involving accretion from the ISM for the INS
\citep[e.g.,][]{w97} have essentially been ruled out as energetically
important: the high velocities ($\sim 200\mbox{ km s}^{-1}$) inferred
for \rxjw\ and \rxjk\ from their proper motions and parallaxes
\citep{kvka02,kvka07} make Bondi-Hoyle accretion (${\dot M}\propto
v^{-3}$) very improbable, especially with the ISM density for \rxjw\
inferred from the H$\alpha$ nebula \citep{vkk01b}.  
%Furthermore, models for the H$\alpha$
%nebula around \rxjw\ \citep{vkk01b} lead to estimates for the ambient
%ISM density that essentially exclude ISM accretion.  
We can then
examine accretion from a fossil fall-back disk \citep{chn00,alpar01}.
Timing observations of \rxjk\ limit the torques from a
disk such that  $\lsim1$\% of the observed X-ray luminosity can come
from frictional heating, and having accretion penetrate the
centrifugal barrier to land on the surface would require an
unreasonably large accretion rate of ${\dot M} \sim
\expnt{2}{17}\mbox{ g s}^{-1}$ \citep[also see][]{lcmpi07}. 

%\paragraph{Magnetic Fields}
Excluding accretion as an energy source, we are left with residual
thermal energy and magnetic fields.  All neutron stars should have
thermal energy powering emission at some level, but only those with
magnetic fields high enough  to decay ($\gsim 10^{14}$~G, generally)
will have significant magnetic luminosity \citep{hk98}.

\citet{hk98} proposed a model where \rxjk\ was heated by magnetic
field decay.  This would help explain the similarity between the
spin-period of \rxjk\ and those of magnetars \cite{wt06}, and it would
also explain why objects like \rxjk\ could be overrepresented in
a local sample.  However, timing
observations of \rxjk\ suggest that the magnetic field was not
ever strong enough to decay significantly \citep{zhc+02,kkvkm02}.

%\paragraph{Rotation \& Cooling}
We are then left with cooling.  The X-ray emission
from the INS is consistent (given the many uncertainties) with that
expected for standard cooling \citep[e.g.,][]{plps04}, and spectral modeling
finds no detectable evidence for non-thermal emission. This is largely
consistent with the observed $\dot E$ and X-ray luminosity, as
discussed above.  Indeed, for most young neutron stars the X-ray
emission that we see is some combination of cooling and rotation
\citep[e.g.,][]{kpzr05}, with the relative fractions varying depending
on source age, $\dot E$, and distance (for the more distant sources,
interstellar absorption tends to remove traces of soft thermal
emission first). \citet{zhc+02} and \citet{kkvkm02} suggested that
\rxjk\ (and the rest of the INS) were in fact rotation-powered pulsars
either without radio emission or where the radio beam does not cross
our line of sight (consistent with the very narrow beams found for
long-period pulsars \citep{ymj99,rankin93,tm98}), but where we do not
actually see any of the non-thermal emission.  This is consistent with
the interpretation of the H$\alpha$ nebula around \rxjw\ as a
bow-shock, which requires an energetic particle wind to come from the
neutron star (presumably as a result of spin-down).  The small
deviations of \rxjk\ from a Rayleigh-Jeans spectrum in the optical may
be a hint of a non-thermal power-law peaking out, or they may just
reflect our lack of understanding of the intrinsic spectrum.  The spin
characteristics of the INS are unusual compared to the bulk of the pulsar
population but are not entirely unheard of --- we will return to how
the INS relate to the pulsar population later.

\section{The Utility of the INS}
Accepting that the INS are nearby, cooling neutron stars with moderate
($\sim 10^{13}$~G) magnetic fields and cold ($kT \lsim 100$~eV)
surfaces, the sources take on importance beyond their contribution to
the local neutron star population.  We can use the INS as ``physics
laboratories,'' probing regimes of physics not accessible
experimentally (or sometimes even theoretically \citep{lp04}).  There
are three areas where we can hope to derive meaningful physical
constraints from or explore novel regimes with the INS: (1) neutron
star radius measurements; (2) neutron star cooling; and (3) strong
magnetic fields.  While measurements of other neutron stars can be
used to address these issues, often providing complementary
constraints \citetext{many explored in these proceedings}, the INS are
particularly well-suited.  This is because they are nearby, young, and
have no detectable non-thermal emission.  Therefore they are bright,
can still constrain cooling models, and have emission that is less
susceptible to arbitrary decomposition than many rotation-powered
pulsars.

The first two physical constraints  come about because of our
relative ignorance about the details of the equation of state (EoS) of
matter at super-nuclear densities and for neutron-rich systems, such
as those one finds in the interiors of neutron stars.  We do not know
some of the basic properties of such matter such as the constituent
particles, and options ranging from basic nuclear matter to pion or
kaon condensates to color-superconducting quark matter are all
possible.  With each of those possibilities comes a range of
predictions for the overall mass and radius of the neutron star, but
one finds that the radius is largely independent of the mass (at least
for masses near the canonical value of $1.4\,M_{\odot}$), so that
a radius measurement has the hope of constraining the physical models
of the neutron star interior \citep{lp04,lp07}.  At the same time, the
detailed predictions for the interior lead to different microphysical
processes that affect the overall cooling of the neutron star, so
knowledge of how neutron stars cool can also be used to constrain the
interior \citep{yp04,plps04}.

To constrain the radii or cooling properties of the INS, there are a
number of observational quantities that we need to understand.  For
cooling, the goal is to place the INS on a plot of luminosity versus
age\footnote{We could also do temperature versus age, although there
are substantial difficulties in converting reliably between effective
temperature and observed temperature \citep{pgw06}.}.  The ages can be
estimated through two methods: either tracing the neutron stars back
to likely birth locations, or through standard pulsar spin-down
assumptions.  While we would expect both of those methods to have
uncertainties, the first is less likely to be systematically wrong.
Indeed, the spin-down ages for \rxjk\ and \rbs\ are 1--2~Myr, compared
to kinematic ages of $<0.7$~Myr measured for those and two other
sources.  This could be a coincidence, but it does mean that we should
be cautious in using the spin-down ages (also see, e.g., \citep{lkg+07}).
The luminosity needs both a distance (measured through astrometry or
estimated from absorption column densities) and an accurate flux
measurement.  However, as discussed in \citet{pgw06}, the influence of
the neutron star's envelope on the emergent spectrum and the cooling
behavior is such that we also need to understand the elemental
abundances, surface temperature distribution, and magnetic field
distribution over the surface (also see \citep[][Yakovlev, these
proceedings]{yp04,pgk07}).  This last item has been the most
difficult, and we will return to it below.  Radius measurements need
essentially the same data, although we do not need an age.  But
accurate distances and flux measurements are just as important, and
again our ignorance of the surface composition etc.\ limit any
possible conclusions \citep{pwl+02,hkc+07}.

Investigating magnetic fields is a more tractable problem.  Instead of
the Quantum Chromodynamics (QCD) in the neutron star interior, where
the current theory leaves many areas unknown \citep{rho00}, the magnetic fields
manifest Quantum Electrodynamics (QED) which is difficult but solvable
in this regime.  Here it is not so much that we use the neutron stars
to constrain physics as we have two-way feedback between the theory
and the observations.  The physical phase of the surface
material can be strongly affected by the magnetic field \citep{ml06},
and the magnetic field also affects the radiation propagating through
it.  The vacuum is polarized, and density gradients can lead to
transitions between the polarization states that have significant
effects on the total spectrum \citep{hl03}.  Again, these are areas
where the fundamentals are known, but there are no terrestrial tests
to the theory in these regimes and hence observations of neutron stars
form the best application of these effects.

\begin{figure}
\includegraphics[width=\hsize]{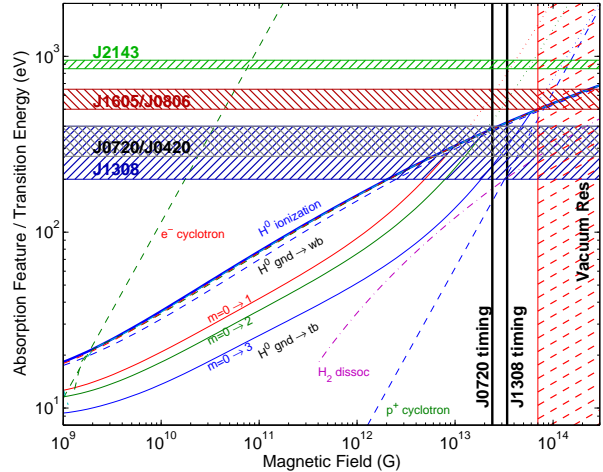}
\caption{Energy versus magnetic field for electron and proton
cyclotron, ground state to tightly bound (tb) or weakly bound (wb)
states in neutral hydrogen, hydrogen ionization, and molecular H$_2$
dissociation \citep{hlpc03,lai01,potekhin98}. For fields above
$\sim\!10^{14}\,$G (hatched region to the right), features may partly
be washed out due to the effects of vacuum resonance mode
conversion \citep{lh03}.  The hatched bands show the absorption feature
energies (corrected for gravitational redshift) for the INS
(Table~\ref{tab:sum}).  We argue \citep{vkk07} that the features in
most sources are due to to neutral hydrogen (at least in part), the
only exception being RX~J2143, which has the highest energy, and for
which proton cyclotron absorption appears more likely.  Qualitatively,
the  field strengths determined by timing for RX~J0720 and
RX~J1308, indicated by the vertical lines, are consistent with our
identifications.  
}
\label{fig:h}
\end{figure}

\section{Open Questions}
%\subsection{Energy Sources}

\subsection{Understanding the Atmospheres}
Initial attempts to model the X-ray spectra of the INS focused on
blackbodies \citep{wal01,pmm+01}, as they gave adequate fits and were
simple.  Including the optical/UV data it was clear that a single
blackbody would not fit the data.  Instead, two blackbodies were used,
where a hot, small surface fit the X-ray data and a cooler, large
surface fit the optical/UV \citep{br02,kvkm+03}.  However, detailed
applications of these models did not work, especially when one
considered the phase-dependent spectral evolution \citep{shhm05,zt06},
and such models are not physical. 

Instead we must consider more realistic models, taking into account
our latest knowledge of the X-ray spectra plus any additional
constraints such as dipole magnetic fields from X-ray timing.  
In interpreting the spectra, a major uncertainty is the composition.
For a single source, this may be difficult to determine uniquely, but
one can hope to make progress by treating the INS as an ensemble:
ideally, it should be possible to understand the features (or lack
thereof) in all INS with a single composition, appealing only to
differences in temperature and magnetic field strength (constrained by
observations where possible), which might lead to different ionization
states being dominant, and possibly the formation of molecules or even
a condensate \citep{lai01,pcs99}.

There are a range of possibilities to explain the overall spectra and
the X-ray absorption features of the INS.  For the absorption, the
simplest models are either proton cyclotron resonances or bound states
of hydrogen, although other species are of course possible (though
possibly more difficult to explain, as gaseous atmospheres composed of
heavier elements appear to be excluded by the lack of large numbers of
features, and any hydrogen should float to the surface).

We \citep{vkk07} have attempted to determine which species and
magnetic field is dominant for each of the INS, based on the energies,
strengths, and widths of the absorption features (Fig.~\ref{fig:h}).
We can find a moderately self-consistent picture, with both proton
cyclotron and hydrogen contributing.  This picture agrees reasonably
well with the magnetic fields inferred from X-ray timing although it
was conceived independently.  However, it is incomplete in at least
two aspects.  First, the evidence for multiple absorption lines
\citep{haberl07,shhm07}  does not fit into the scenario easily.
Second, the evolution in the spectrum of \rxjk\ is also not
understood.  See \citet{vkk07} for more discussion of both of these
aspects, as well as more general issues.

In order to fit both the X-ray and optical data simultaneously,
\citet{mzh03} and \citet{tzd04} considered models that have condensed,
blackbody-like surfaces, where \citet{mzh03} also includes a thin
layer of hydrogen that is optically thin at X-ray energies but not at
optical.  Getting a condensed surface is possible for this range of
magnetic field and temperature \citep{ml06,ml07b}, although it depends
on the composition and the magnetic field distribution.  Including the
thin layer of hydrogen  and varying the magnetic field allow
for tuning of the optical excess to match observations.  Indeed,
\citet{hkc+07} have updated this model to include partially-ionized
hydrogen and found a reasonable fit to the phase-averaged spectrum of
\rxjw\ as well as the amplitude and shape of its pulsations
\citep{ho07}, making this a promising avenue of investigation.
However, we must still consider a number of issues addressing the
physical reality of these models: How did such a thin layer form?  Is
it stable?  Will it persist?  Why are the layers of roughly the same
thicknesses for different sources?

\subsection{Evolution of \rxjk}
While the initial observations of \rxjk\ showed a smooth,
blackbody-like X-ray spectrum \citep{pmm+01,kvkm+03}, later
observations showed evidence for a phase-dependent absorption feature
\citep{hztb04}.  This discrepancy was first ascribed to increased
sensitivity and better calibration of later observations, but
\citet{dvvmv04} realized that \rxjk\ was actually evolving over the
course of months.  The spectrum was changing, with the blackbody
getting harder at the same time as low-energy absorption appeared,
such that the flux was relatively constant.
Simultaneously, the pulse profile evolved to increase the pulsed
fraction \citep{vdvmv04}.  The change happened over the course of
several years, but the majority of the variation occurred over the time
span of a few months, from May to October of 2003.

These observations were very puzzling, as \rxjk\ was not expected to
vary on such time scales.  Two sets of models were considered to
explain the data \citep{dvvmv04}: either the intrinsic spectrum of
\rxjk\ changed, or the angle at which we view the object change.  They
conclude that the second model is more likely, as no obvious physical
motivation could be found for the first.  Instead, free precession
\citep{link03} could be operating.  \citet{htdv+06} find further
support for this model by fitting additional data and observing that
the spectral changes seem to be reversing themselves after several
years; they suggest a timescale of $\sim 7$~yr for the precession,
although note that this was with only $\sim 5$~yr of spectral data, so
the timescale was not well constrained.  If true, this would have
important implications for our study of the superfluid states in the
object's interior \citep{link03}.

However, other models are still possible.  We have investigated the
spectral data and the timing data together \citep{vkkpm07}, and find
evidence that the changes in the spectrum appear more impulsive than
periodic and were accompanied by a simultaneous jump in the spin
frequency, with a fractional increase of $ \expnt{5}{-8}$.  Instead of
precession, we interpret this as evidence for an sudden change on the
neutron star surface accompanied by a simultaneous torque, followed by
a slow relaxation back to the original spectrum.  We have considered a
number of mechanisms, from typical pulsar ``glitches'' that would
release energy in the neutron star interior, to reconfiguration of the
magnetic field leading to changes in both the spin-down and in the
temperature/emergent spectrum of the surface, to accretion of small
amounts of material onto the neutron star surface that would both
torque the star and change the composition.  No mechanism is fully
satisfactory, but we are attempting to include constraints from
modeling the phase-dependence of the spectrum to try to understand the
situation better \citetext{Mori~et~al., in prep}.  We are also looking
for similar behavior in other sources, but so far they seem stable
\citetext{Airhart~et~al., these proceedings}.

\subsection{Relation to the Pulsar Population}
If we consider that the INS are moderately young neutron stars with
$\sim 10^{13}$~G magnetic fields, we must ask what separates them from
the bulk of the rotation-powered pulsar population and what their
relation to that population is.  To gain insight, we can compare with
two groups of objects.  First, the other pulsars found with comparable
count rates ($>0.05\mbox{ s}^{-1}$) in the RASS, and second, the known
pulsars with $B\gsim 10^{13}$~G.

%\paragraph{\rosat\ Sources} 
As highlighted by its use is discovering the INS, the RASS is an
efficient and relatively unbiased way to find young neutron stars.
All neutron stars should shine in soft X-rays, either via cooling
radiation or non-thermal processes.  Taking only the thermal emission,
which should persist as a baseline, the effects of beaming are quite
minor (compared to radio pulsar searches).  Aside from the 7 INS,
there are 8 other Galactic pulsars with comparable count-rates
\citep[based on][]{bt97,pccm02,kaplan04}.  These include many of the
best-known pulsars: the Crab pulsar, the Vela pulsar, PSR~B0656+14,
Geminga, and PSRs~1055$-$52, J0437$-$4715, B1951+32, and J0538+2817
(in order of decreasing count-rate; see squares in
Fig.~\ref{fig:ppdot}).  Of these, the Crab is much younger and more
energetic than the others but is significantly more distant, and
PSR~J0437$-$4715 is an old recycled pulsar.  The remaining objects are
moderately young and nearby, much like the INS.  However, the INS all
have periods $>3$~s (known for 6 of 7), while the other pulsars all
have periods $<0.4$~s.  Not only that, but the other pulsars all have
detectable non-thermal X-ray emission.  In fact, if we look at the
characteristics that define the INS --- young, nearby, X-ray bright,
long periods --- we cannot find any other pulsars in the ATNF catalog
\citep{mhth05} that match them.  The INS are obviously then a
significant subpopulation of objects, representing as many as half of
the observed neutron stars within $\sim 500$~pc.  The number of
detected INS compared to the number of young pulsars in the same
volume implies a very large total number of similar sources in the
Galaxy. There have been a number of population estimates
\citep[e.g.,][]{ptp+05,ptp06} that try (with moderate success) to
explore this quantitatively, but they are limited by the small number
of sources.  However, we must still explain why there are so many INS
with similar periods (and presumably ages and magnetic fields) in such
a small volume.  The idea of \citet{hk98}, that magnetic-field decay
could lead to a preponderance of nearby old magnetars, is certainly
worth examining.  While as discussed above the INS do not seem to have
been true magnetars in the past, these conclusions used simple models
for field decay (something still not understood in detail), and they
assumed that the internal field was comparable to the dipolar
component.  If there are significant toroidal components to the field,
these may be strong enough to decay but would not affect spin-down.  Therefore, the INS could have some contribution to
their X-ray luminosity from magnetic field decay, or magnetic fields
could alter standard cooling behavior if they are strong enough to
influence the heat conduction through to the surface \citep{pgk07}.

%\paragraph{High-$B$ Radio Pulsars} 
If we the look further afield for objects that more closely resemble
the INS, we find the so-called high-$B$ radio pulsars (HBPSRs).  These
sources are largely emerging in the last seven years and their
population is probably not complete, but we see examples with dipolar
fields of several $\times 10^{13}$~G, almost up to $10^{14}$~G
\citep{ckl+00,mhl+02,msk+03}.  Some of the magnetic fields are
actually higher than those of the INS, but the distribution of sources
with respect to $B_{\rm dip}$ (Fig.~\ref{fig:ppdot}) seems mostly
continuous leading up to and past \rxjk\ and \rbs.  The HBPSRs are
younger than the INS, but tend to be much more distant and have
higher values of $\dot E$/X-ray luminosity (when detected).  This last
fact could possibly be explained just through the usual models for
pulsar evolution: assuming a constant magnetic field, then $\dot E$
evolves as $B_{\rm dip}^{-2}t^{-2}$, so the difference of a factor of
roughly $10^2$ in age would correspond to $10^4$ in $\dot E$.  This
would then give values consistent with those for the INS.  At the
same time, as $\dot E$ dropped the dominant source of X-ray emission
became the residual thermal emission that we see from the INS instead
of the power-law emission that we see from the HBPSRs (and from other
active pulsars).

The INS may then represent a population of evolved HBPSRs
\citep{zhc+02,kkvkm02}.  Previous analyses of pulsar populations have
usually not required pulsars beyond magnetic fields of $10^{13}$~G or
so \citep[e.g.,][]{narayan87,no90,acc02}, but it now seems apparent
that the true distribution extends further in significant numbers (a
conclusion also becoming apparent just from recent pulsar surveys;
\citep{vml+04,fgk06}).  

As discussed above, there are no detections of radio emission (pulsed
or continuous) for the INS.  The flux limits are reasonably low, and
coupled with the small distances to the INS the luminosity limits are
orders of magnitude below the luminosities of the faintest known radio
pulsars (such as PSR~J0205+6449 in 3C~58; \citep{csl+02}).  The
non-detection of radio emission from the INS could be explained by the
very narrow radio beams found for long-period pulsars/HBPSRs ($\lsim
1$\% \citep{msk+03,ymj99}); this has led to large uncertainties in the
predicted number of long-period sources \citep[e.g.,][]{narayan87}.
At the same time, the rapid evolution of the HBPSRs across the $P-\dot
P$ plane may drive them across the ``death line'' and terminate radio
activity quickly, so that the objects we see may no longer be radio
sources at all.  We also look to the sporadically emitting RRATs, some
of which seem to have similar spin characteristics to the INS and
which may be a related population \citep{ptp06}.

%\section{Conclusions}

\section{Discovering New Sources}
With such a limited sample, the INS are invaluable for individual
investigations but have only limited use as a population.  In
addition, each object has its own peculiarities and pathologies, which
make comparison with the rest of the sample difficult.  For instance,
of the two brightest objects, \rxjw\ lacks an X-ray absorption feature
but has an H$\alpha$ nebula, while \rxjk\ varies.

To help remedy this situation, for several years intensive efforts
have been underway to identify new isolated neutron stars \citep[][
Pires et~al.\ in these proceedings]{rfbm03,aam+06}.  This is a
difficult and time-consuming task, given the need to cross-identify
X-ray sources (which often have poorly known positions) with optical
catalogs, rejecting those sources that have plausible counterparts
\citep[e.g.,][]{rbpl00}.

Currently, the majority of these efforts have yet to produce
significant results, but the first few good candidates are emerging.
Using the \textit{SWIFT} satellite, \citet{fox04} is improving the
positions of \rosat\ sources, which facilitates the search.
Based on this, \citet{rfs07} have identified a source which appears to be a neutron
star, although not necessarily of the same type (young, thermally
emitting) as the INS.  Pires~et~al.\ [these proceedings] may have
identified another from \xmm\ data, although given how faint and
absorbed that object is likely to be, confirming it as a neutron star
with deep optical observations will be very difficult, and X-ray data
will only yield limited information.  Even so, as more of these
objects come to light they can help us construct more reliable
population estimates for the INS and establish their relation to the
greater neutron star population.

\begin{theacknowledgments}
I thank Marten van~Kerkwijk for many stimulating discussions.
\end{theacknowledgments}

\bibliographystyle{aipproc}   % if natbib is available


%\bibliography{ins}

\begin{thebibliography}{113}
\expandafter\ifx\csname natexlab\endcsname\relax\def\natexlab#1{#1}\fi
\providecommand{\enquote}[1]{``#1''}
\expandafter\ifx\csname url\endcsname\relax
  \def\url#1{\texttt{#1}}\fi
\expandafter\ifx\csname urlprefix\endcsname\relax\def\urlprefix{URL }\fi
\providecommand{\eprint}[2][]{\url{#2}}

\bibitem[{Vranesevic} et~al.(2004)]{vml+04}
N.~{Vranesevic}, et~al., \emph{\apjl} \textbf{617}, L139 (2004).

\bibitem[{Faucher-Gigu{\`e}re} and {Kaspi}(2006)]{fgk06}
C.-A. {Faucher-Gigu{\`e}re}, and V.~M. {Kaspi}, \emph{\apj} \textbf{643}, 332
  (2006).

\bibitem[{Ostriker} et~al.(1970)]{ors70}
J.~P. {Ostriker}, M.~J. {Rees}, and J.~{Silk}, \emph{\aplett} \textbf{6}, L179
  (1970).

\bibitem[{Treves} and {Colpi}(1991)]{tc91}
A.~{Treves}, and M.~{Colpi}, \emph{\aap} \textbf{241}, 107 (1991).

\bibitem[{Blaes} and {Madau}(1993)]{bm93}
O.~{Blaes}, and P.~{Madau}, \emph{\apj} \textbf{403}, 690 (1993).

\bibitem[{Treves} et~al.(2000)]{ttzc00}
A.~{Treves}, R.~{Turolla}, S.~{Zane}, and M.~{Colpi}, \emph{\pasp}
  \textbf{112}, 297 (2000).

\bibitem[{Hertz} and {Grindlay}(1988)]{hg88}
P.~{Hertz}, and J.~E. {Grindlay}, \emph{\aj} \textbf{96}, 233 (1988).

\bibitem[{Hobbs} et~al.(2005)]{hllk05}
G.~{Hobbs}, D.~R. {Lorimer}, A.~G. {Lyne}, and M.~{Kramer}, \emph{\mnras}
  \textbf{360}, 974 (2005).

\bibitem[{Neuh\"{a}user} and {Tr\"{u}mper}(1999)]{nt99}
R.~{Neuh\"{a}user}, and J.~E. {Tr\"{u}mper}, \emph{\aap} \textbf{343}, 151
  (1999).

\bibitem[{Illarionov} and {Sunyaev}(1975)]{is75}
A.~F. {Illarionov}, and R.~A. {Sunyaev}, \emph{\aap} \textbf{39}, 185 (1975).

\bibitem[{Perna} et~al.(2003)]{pnr+03}
R.~{Perna}, R.~{Narayan}, G.~{Rybicki}, L.~{Stella}, and A.~{Treves},
  \emph{\apj} \textbf{594}, 936 (2003).

\bibitem[{Walter} et~al.(1996)]{wwn96}
F.~M. {Walter}, S.~J. {Wolk}, and R.~{Neuh\"{a}user}, \emph{\nat} \textbf{379},
  233 (1996).

\bibitem[{Knude} and {H{\o}g}(1998)]{kh98}
J.~{Knude}, and E.~{H{\o}g}, \emph{\aap} \textbf{338}, 897 (1998).

\bibitem[{van Kerkwijk} and {Kulkarni}(2001{\natexlab{a}})]{vkk01}
M.~H. {van Kerkwijk}, and S.~R. {Kulkarni}, \emph{\aap} \textbf{378}, 986
  (2001{\natexlab{a}}).

\bibitem[{Kaplan} et~al.(2002{\natexlab{a}})]{kvka02}
D.~L. {Kaplan}, M.~H. {van Kerkwijk}, and J.~{Anderson}, \emph{\apj}
  \textbf{571}, 447 (2002{\natexlab{a}}).

\bibitem[{Walter} and {Lattimer}(2002)]{wl02}
F.~M. {Walter}, and J.~M. {Lattimer}, \emph{\apjl} \textbf{576}, L145 (2002).

\bibitem[{Walter} and {Matthews}(1997)]{wm97}
F.~M. {Walter}, and L.~D. {Matthews}, \emph{\nat} \textbf{389}, 358 (1997).

\bibitem[{Burwitz} et~al.(2003)]{bhn+03}
V.~{Burwitz}, F.~{Haberl}, R.~{Neuh{\" a}user}, P.~{Predehl}, J.~{Tr{\"
  u}mper}, and V.~E. {Zavlin}, \emph{\aap} \textbf{399}, 1109 (2003).

\bibitem[{Tiengo} and {Mereghetti}(2007)]{tm07}
A.~{Tiengo}, and S.~{Mereghetti}, \emph{\apjl} \textbf{657}, L101 (2007).

\bibitem[{van Kerkwijk} and {Kaplan}(2007)]{vkk07}
M.~H. {van Kerkwijk}, and D.~L. {Kaplan}, \emph{\apss} \textbf{308}, 191
  (2007).

\bibitem[{Kaplan} et~al.(2003{\natexlab{a}})]{kvkm+03}
D.~L. {Kaplan}, et~al., \emph{\apj} \textbf{590}, 1008 (2003{\natexlab{a}}).

\bibitem[{Motch} et~al.(2003)]{mzh03}
C.~{Motch}, V.~E. {Zavlin}, and F.~{Haberl}, \emph{\aap} \textbf{408}, 323
  (2003).

\bibitem[{Kaplan} and {van Kerkwijk}(2005{\natexlab{a}})]{kvk05}
D.~L. {Kaplan}, and M.~H. {van Kerkwijk}, \emph{\apjl} \textbf{628}, L45
  (2005{\natexlab{a}}).

\bibitem[{Kaplan} et~al.(2007)]{kvka07}
D.~L. {Kaplan}, M.~H. {van Kerkwijk}, and J.~{Anderson}, \emph{\apj}
  \textbf{660}, 1428 (2007).

\bibitem[{van Kerkwijk} et~al.(2007)]{vkkpm07}
M.~H. {van Kerkwijk}, D.~L. {Kaplan}, G.~G. {Pavlov}, and K.~{Mori},
  \emph{\apjl} \textbf{659}, L149 (2007).

\bibitem[{Haberl} et~al.(2004{\natexlab{a}})]{hztb04}
F.~{Haberl}, V.~E. {Zavlin}, J.~{Tr{\" u}mper}, and V.~{Burwitz}, \emph{\aap}
  \textbf{419}, 1077 (2004{\natexlab{a}}).

\bibitem[{Kaplan} et~al.(2003{\natexlab{b}})]{kkvk03}
D.~L. {Kaplan}, S.~R. {Kulkarni}, and M.~H. {van Kerkwijk}, \emph{\apjl}
  \textbf{588}, L33 (2003{\natexlab{b}}).

\bibitem[{van Kerkwijk} et~al.(2004)]{vkkd+04}
M.~H. {van Kerkwijk}, D.~L. {Kaplan}, M.~{Durant}, S.~R. {Kulkarni}, and
  F.~{Paerels}, \emph{\apj} \textbf{608}, 432 (2004).

\bibitem[{Haberl}(2007)]{haberl07}
F.~{Haberl}, \emph{\apss} \textbf{308}, 181 (2007).

\bibitem[{Motch} et~al.(2005)]{msh+05}
C.~{Motch}, et~al., \emph{\aap} \textbf{429}, 257 (2005).

\bibitem[{Zane} et~al.(2006)]{zdlmt06}
S.~{Zane}, A.~{de Luca}, R.~P. {Mignani}, and R.~{Turolla}, \emph{\aap}
  \textbf{457}, 619 (2006).

\bibitem[{Kaplan} et~al.(2002{\natexlab{b}})]{kkvk02}
D.~L. {Kaplan}, S.~R. {Kulkarni}, and M.~H. {van Kerkwijk}, \emph{\apjl}
  \textbf{579}, L29 (2002{\natexlab{b}}).

\bibitem[{Kaplan} and {van Kerkwijk}(2005{\natexlab{b}})]{kvk05b}
D.~L. {Kaplan}, and M.~H. {van Kerkwijk}, \emph{\apjl} \textbf{635}, L65
  (2005{\natexlab{b}}).

\bibitem[{Haberl} et~al.(2003)]{hsh+03}
F.~{Haberl}, A.~D. {Schwope}, V.~{Hambaryan}, G.~{Hasinger}, and C.~{Motch},
  \emph{\aap} \textbf{403}, L19 (2003).

\bibitem[{Schwope} et~al.(2005)]{shhm05}
A.~D. {Schwope}, V.~{Hambaryan}, F.~{Haberl}, and C.~{Motch}, \emph{\aap}
  \textbf{441}, 597 (2005).

\bibitem[{Schwope} et~al.(2007)]{shhm07}
A.~D. {Schwope}, V.~{Hambaryan}, F.~{Haberl}, and C.~{Motch}, \emph{\apss}
  \textbf{308}, 619 (2007).

\bibitem[{Zampieri} et~al.(2001)]{zct+01}
L.~{Zampieri}, et~al., \emph{\aap} \textbf{378}, L5 (2001).

\bibitem[{Zane} et~al.(2005)]{zct+05}
S.~{Zane}, et~al., \emph{\apj} \textbf{627}, 397 (2005).

\bibitem[{Rea} et~al.(2007)]{rtj+07}
N.~{Rea}, et~al., \emph{\mnras} \textbf{379}, 1484 (2007).

\bibitem[{Haberl} et~al.(2004{\natexlab{b}})]{hmz+04}
F.~{Haberl}, et~al., \emph{\aap} \textbf{424}, 635 (2004{\natexlab{b}}).

\bibitem[{Posselt} et~al.(2007)]{pph+07}
B.~{Posselt}, S.~B. {Popov}, F.~{Haberl}, J.~{Tr{\"u}mper}, R.~{Turolla}, and
  R.~{Neuh{\"a}user}, \emph{\apss} \textbf{308}, 171 (2007).

\bibitem[{Haberl} et~al.(1997)]{hmb+97}
F.~{Haberl}, C.~{Motch}, D.~A.~H. {Buckley}, F.-J. {Zickgraf}, and
  W.~{Pietsch}, \emph{\aap} \textbf{326}, 662 (1997).

\bibitem[{Haberl} et~al.(1998)]{hmp98}
F.~{Haberl}, C.~{Motch}, and W.~{Pietsch}, \emph{Astronomische Nachrichten}
  \textbf{319}, 97 (1998).

\bibitem[{Schwope} et~al.(1999)]{shs+99}
A.~D. {Schwope}, G.~{Hasinger}, R.~{Schwarz}, F.~{Haberl}, and M.~{Schmidt},
  \emph{\aap} \textbf{341}, L51 (1999).

\bibitem[{Motch} et~al.(1999)]{mhz+99}
C.~{Motch}, F.~{Haberl}, F.-J. {Zickgraf}, G.~{Hasinger}, and A.~D. {Schwope},
  \emph{\aap} \textbf{351}, 177 (1999).

\bibitem[{Haberl} et~al.(1999)]{hpm99}
F.~{Haberl}, W.~{Pietsch}, and C.~{Motch}, \emph{\aap} \textbf{351}, L53
  (1999).

\bibitem[{Rutledge} et~al.(2003)]{rfbm03}
R.~E. {Rutledge}, D.~W. {Fox}, M.~{Bogosavljevic}, and A.~{Mahabal},
  \emph{\apj} \textbf{598}, 458 (2003).

\bibitem[{Voges} et~al.(1999)]{rbs}
W.~{Voges}, et~al., \emph{\aap} \textbf{349}, 389 (1999).

\bibitem[{Popov} et~al.(2005)]{ptp+05}
S.~B. {Popov}, R.~{Turolla}, M.~E. {Prokhorov}, M.~{Colpi}, and A.~{Treves},
  \emph{\apss} \textbf{299}, 117 (2005).

\bibitem[{Durant} and {van Kerkwijk}(2006)]{dvk06}
M.~{Durant}, and M.~H. {van Kerkwijk}, \emph{\apj} \textbf{650}, 1082 (2006).

\bibitem[{Lorimer} and {Kramer}(2004)]{lk04}
D.~R. {Lorimer}, and M.~{Kramer}, \emph{{Handbook of Pulsar Astronomy}},
  Cambridge University Press, Cambridge, UK, 2004.

\bibitem[{Kulkarni} and {van~Kerkwijk}(1998)]{kvk98}
S.~R. {Kulkarni}, and M.~H. {van~Kerkwijk}, \emph{\apjl} \textbf{507}, L49
  (1998).

\bibitem[{Motch} and {Haberl}(1998)]{mh98}
C.~{Motch}, and F.~{Haberl}, \emph{\aap} \textbf{333}, L59 (1998).

\bibitem[{van Kerkwijk} and {Kulkarni}(2001{\natexlab{b}})]{vkk01b}
M.~H. {van Kerkwijk}, and S.~R. {Kulkarni}, \emph{\aap} \textbf{380}, 221
  (2001{\natexlab{b}}).

\bibitem[{Chatterjee} and {Cordes}(2002)]{cc02}
S.~{Chatterjee}, and J.~M. {Cordes}, \emph{\apj} \textbf{575}, 407 (2002).

\bibitem[{Walter}(2001)]{wal01}
F.~M. {Walter}, \emph{\apj} \textbf{549}, 433 (2001).

\bibitem[{Johnston}(2003)]{johnston03}
S.~{Johnston}, \emph{\mnras} \textbf{340}, L43 (2003).

\bibitem[{McLaughlin} et~al.(2006)]{mll+06}
M.~A. {McLaughlin}, et~al., \emph{\nat} \textbf{439}, 817 (2006).

\bibitem[{Lo Curto} et~al.(2007)]{lcmpi07}
G.~{Lo Curto}, R.~P. {Mignani}, R.~{Perna}, and G.~L. {Israel}, \emph{\aap}
  \textbf{in press} (2007), \eprint{arXiv:astro-ph/0707.0516}.

\bibitem[{Becker} and {Tr\"{u}mper}(1997)]{bt97}
W.~{Becker}, and J.~{Tr\"{u}mper}, \emph{\aap} \textbf{326}, 682 (1997).

\bibitem[{Possenti} et~al.(2002)]{pccm02}
A.~{Possenti}, R.~{Cerutti}, M.~{Colpi}, and S.~{Mereghetti}, \emph{\aap}
  \textbf{387}, 993 (2002).

\bibitem[{Wang}(1997)]{w97}
J.~C.~L. {Wang}, \emph{\apjl} \textbf{486}, L119 (1997).

\bibitem[{Alpar}(2001)]{alpar01}
M.~A. {Alpar}, \emph{\apj} \textbf{554}, 1245 (2001).

\bibitem[Chatterjee et~al.(2000)]{chn00}
P.~Chatterjee, L.~Hernquist, and R.~Narayan, \emph{\apj} \textbf{534}, 373
  (2000).

\bibitem[{Manchester} et~al.(2005)]{mhth05}
R.~N. {Manchester}, G.~B. {Hobbs}, A.~{Teoh}, and M.~{Hobbs}, \emph{\aj}
  \textbf{129}, 1993 (2005).

\bibitem[{Popov} et~al.(2006)]{ptp06}
S.~B. {Popov}, R.~{Turolla}, and A.~{Possenti}, \emph{\mnras} \textbf{369}, L23
  (2006).

\bibitem[{Heyl} and {Kulkarni}(1998)]{hk98}
J.~S. {Heyl}, and S.~R. {Kulkarni}, \emph{\apjl} \textbf{506}, L61 (1998).

\bibitem[{Woods} and {Thompson}(2006)]{wt06}
P.~M. {Woods}, and C.~{Thompson}, \enquote{{Soft gamma repeaters and anomalous
  X-ray pulsars: magnetar candidates},} in \emph{Compact stellar X-ray
  sources}, edited by W.~{Lewin}, and M.~{van der Klis}, Cambridge University
  Press, Cambridge, UK, 2006, p. 547.

\bibitem[{Zane} et~al.(2002)]{zhc+02}
S.~{Zane}, et~al., \emph{\mnras} \textbf{334}, 345 (2002).

\bibitem[{Kaplan} et~al.(2002{\natexlab{c}})]{kkvkm02}
D.~L. {Kaplan}, S.~R. {Kulkarni}, M.~H. {van Kerkwijk}, and H.~L. {Marshall},
  \emph{\apjl} \textbf{570}, L79 (2002{\natexlab{c}}).

\bibitem[{Page} et~al.(2004)]{plps04}
D.~{Page}, J.~M. {Lattimer}, M.~{Prakash}, and A.~W. {Steiner}, \emph{\apjs}
  \textbf{155}, 623 (2004).

\bibitem[{Kargaltsev} et~al.(2005)]{kpzr05}
O.~Y. {Kargaltsev}, G.~G. {Pavlov}, V.~E. {Zavlin}, and R.~W. {Romani},
  \emph{\apj} \textbf{625}, 307 (2005).

\bibitem[{Young} et~al.(1999)]{ymj99}
M.~D. {Young}, R.~N. {Manchester}, and S.~{Johnston}, \emph{\nat} \textbf{400},
  848 (1999).

\bibitem[{Rankin}(1993)]{rankin93}
J.~M. {Rankin}, \emph{\apj} \textbf{405}, 285 (1993).

\bibitem[{Tauris} and {Manchester}(1998)]{tm98}
T.~M. {Tauris}, and R.~N. {Manchester}, \emph{\mnras} \textbf{298}, 625 (1998).

\bibitem[{Lattimer} and {Prakash}(2004)]{lp04}
J.~M. {Lattimer}, and M.~{Prakash}, \emph{Science} \textbf{304}, 536 (2004).

\bibitem[{Lattimer} and {Prakash}(2007)]{lp07}
J.~M. {Lattimer}, and M.~{Prakash}, \emph{\physrep} \textbf{442}, 109 (2007).

\bibitem[{Yakovlev} and {Pethick}(2004)]{yp04}
D.~G. {Yakovlev}, and C.~J. {Pethick}, \emph{\araa} \textbf{42}, 169 (2004).

\bibitem[{Page} et~al.(2006)]{pgw06}
D.~{Page}, U.~{Geppert}, and F.~{Weber}, \emph{Nuclear Physics A} \textbf{777},
  497 (2006).

\bibitem[{Livingstone} et~al.(2007)]{lkg+07}
M.~A. {Livingstone}, et~al., \emph{\apss} \textbf{308}, 317 (2007).

\bibitem[{Page} et~al.(2007)]{pgk07}
D.~{Page}, U.~{Geppert}, and M.~{K{\"u}ker}, \emph{\apss} \textbf{308}, 403
  (2007).

\bibitem[{Pons} et~al.(2002)]{pwl+02}
J.~A. {Pons}, et~al., \emph{\apj} \textbf{564}, 981 (2002).

\bibitem[{Ho} et~al.(2007)]{hkc+07}
W.~C.~G. {Ho}, D.~L. {Kaplan}, P.~{Chang}, M.~{van Adelsberg}, and A.~Y.
  {Potekhin}, \emph{\mnras} \textbf{375}, 821 (2007).

\bibitem[Rho(2000)]{rho00}
M.~Rho  (2000), \eprint{arXiv:nucl-th/0007073}.

\bibitem[{Medin} and {Lai}(2006)]{ml06}
Z.~{Medin}, and D.~{Lai}, \emph{Phys.\ Rev.\ A} \textbf{74}, 062508 (2006).

\bibitem[{Ho} and {Lai}(2003)]{hl03}
W.~C.~G. {Ho}, and D.~{Lai}, \emph{\mnras} \textbf{338}, 233 (2003).

\bibitem[{Lai}(2001)]{lai01}
D.~{Lai}, \emph{Rev. Modern Physics} \textbf{73}, 629 (2001).

\bibitem[{Ho} et~al.(2003)]{hlpc03}
W.~C.~G. {Ho}, D.~{Lai}, A.~Y. {Potekhin}, and G.~{Chabrier}, \emph{\apj}
  \textbf{599}, 1293 (2003).

\bibitem[{Potekhin}(1998)]{potekhin98}
A.~Y. {Potekhin}, \emph{Journal of Physics B Atomic Molecular Physics}
  \textbf{31}, 49 (1998).

\bibitem[{Lai} and {Ho}(2003)]{lh03}
D.~{Lai}, and W.~C.~G. {Ho}, \emph{\apj} \textbf{588}, 962 (2003).

\bibitem[{Paerels} et~al.(2001)]{pmm+01}
F.~{Paerels}, et~al., \emph{\aap} \textbf{365}, L298 (2001).

\bibitem[{Braje} and {Romani}(2002)]{br02}
T.~M. {Braje}, and R.~W. {Romani}, \emph{\apj} \textbf{580}, 1043 (2002).

\bibitem[{Zane} and {Turolla}(2006)]{zt06}
S.~{Zane}, and R.~{Turolla}, \emph{\mnras} \textbf{366}, 727 (2006).

\bibitem[{Potekhin} et~al.(1999)]{pcs99}
A.~Y. {Potekhin}, G.~{Chabrier}, and Y.~A. {Shibanov}, \emph{\pre} \textbf{60},
  2193 (1999).

\bibitem[{Turolla} et~al.(2004)]{tzd04}
R.~{Turolla}, S.~{Zane}, and J.~J. {Drake}, \emph{\apj} \textbf{603}, 265
  (2004).

\bibitem[{Medin} and {Lai}(2007)]{ml07b}
Z.~{Medin}, and D.~{Lai}, \emph{\mnras} \textbf{submitted} (2007),
  \eprint{arXiv:astro-ph/0708.3863}.

\bibitem[{Ho}(2007)]{ho07}
W.~C.~G. {Ho}, \emph{\mnras} \textbf{380}, 71 (2007).

\bibitem[{de Vries} et~al.(2004)]{dvvmv04}
C.~P. {de Vries}, J.~{Vink}, M.~{M{\' e}ndez}, and F.~{Verbunt}, \emph{\aap}
  \textbf{415}, L31 (2004).

\bibitem[{Vink} et~al.(2004)]{vdvmv04}
J.~{Vink}, C.~P. {de Vries}, M.~{M{\' e}ndez}, and F.~{Verbunt}, \emph{\apjl}
  \textbf{609}, L75 (2004).

\bibitem[{Link}(2003)]{link03}
B.~{Link}, \enquote{{Precession of Isolated Neutron Stars},} in \emph{Radio
  Pulsars}, edited by M.~{Bailes}, D.~J. {Nice}, and S.~E. {Thorsett},
  Astronomical Society of the Pacific, San Francisco, 2003, vol. 302, p. 241.

\bibitem[{Haberl} et~al.(2006)]{htdv+06}
F.~{Haberl}, R.~{Turolla}, C.~P. {de Vries}, S.~{Zane}, J.~{Vink},
  M.~{M{\'e}ndez}, and F.~{Verbunt}, \emph{\aap} \textbf{451}, L17 (2006).

\bibitem[{Kaplan}(2004)]{kaplan04}
D.~L. {Kaplan}, \emph{Ph.D.~Thesis} \textbf{California Institute of Technology}
  (2004).

\bibitem[{Camilo} et~al.(2000)]{ckl+00}
F.~{Camilo}, et~al., \emph{\apj} \textbf{541}, 367 (2000).

\bibitem[{Morris} et~al.(2002)]{mhl+02}
D.~J. {Morris}, et~al., \emph{\mnras} \textbf{335}, 275 (2002).

\bibitem[{McLaughlin} et~al.(2003)]{msk+03}
M.~A. {McLaughlin}, et~al., \emph{\apjl} \textbf{591}, L135 (2003).

\bibitem[{Arzoumanian} et~al.(2002)]{acc02}
Z.~{Arzoumanian}, D.~F. {Chernoff}, and J.~M. {Cordes}, \emph{\apj}
  \textbf{568}, 289 (2002).

\bibitem[{Narayan} and {Ostriker}(1990)]{no90}
R.~{Narayan}, and J.~P. {Ostriker}, \emph{\apj} \textbf{352}, 222 (1990).

\bibitem[{Narayan}(1987)]{narayan87}
R.~{Narayan}, \emph{\apj} \textbf{319}, 162 (1987).

\bibitem[{Camilo} et~al.(2002)]{csl+02}
F.~{Camilo}, et~al., \emph{\apjl} \textbf{571}, L41 (2002).

\bibitem[{Ag{\"u}eros} et~al.(2006)]{aam+06}
M.~A. {Ag{\"u}eros}, et~al., \emph{\aj} \textbf{131}, 1740 (2006).

\bibitem[{Rutledge} et~al.(2000)]{rbpl00}
R.~E. {Rutledge}, R.~J. {Brunner}, T.~A. {Prince}, and C.~{Lonsdale},
  \emph{\apjs} \textbf{131}, 335 (2000).

\bibitem[Fox(2004)]{fox04}
D.~B. Fox  (2004), \eprint{arXiv:astro-ph/0403261}.

\bibitem[{Rutledge} et~al.(2007)]{rfs07}
R.~E. {Rutledge}, D.~B. {Fox}, and A.~H. {Shevchuk}, \emph{\apj} \textbf{in
  press} (2007), \eprint{arXiv:astro-ph/0705.1011}.

\end{thebibliography}

\end{document}